\documentstyle[aps,prd,eqsecnum]{revtex}
\input psfig.sty

\begin{document}
\title{A Test of a Test for Chaos}
\author{John D. Barrow and Janna Levin}
\address{DAMTP, Centre for Mathematical Sciences, Cambridge University,
Wilberforce Rd., Cambridge CB3 0WA, UK }


\twocolumn[\hsize\textwidth\columnwidth\hsize\csname
           @twocolumnfalse\endcsname

\maketitle
\widetext

\begin{abstract}
A simple new binary test for chaos has been proposed by Gottwald and
Melbourne.  We apply this test successfully to the Henon-Heiles and Lorenz
systems, demonstrating its applicability to conservative systems, as
well as dissipative systems. The binary test is effective for highly chaotic
Hamiltonian systems and orbits on a strange attractor and is
particularly useful as a marker of the transition from regularity to
chaos. However, we find it is not able to detect more subtle instances
of transient chaos.
\end{abstract}

\medskip
\noindent{PACS 05.45.-a,82.40.Bj}
\medskip
]

\narrowtext

\section{Introduction}

A promising new binary test for chaos has been proposed by Gottwald and
Melbourne \cite{gm}. The test returns 0 for a non-chaotic system and 1 for a
chaotic system. It does not offer any quantitative information and so cannot
quantify the degree of irregularity in the way a Lyapunov exponent might.
However, it can provide a quick and useful diagnostic for chaos.

Consider a base dynamical system ${\bf \dot{x}}={\bf f}({\bf x})$ of any
dimension with orbits ${\bf x}(t)$. The proposed method to determine if it
is chaotic is based on a Euclidean ${\bf E}(2)$ group extension of the
underlying base dynamics \cite{{g1},{g2}}.  The dynamics is explicitly
extended to include two new variables $(p,q)$ defined through 
\begin{eqnarray}
\dot{p} &=&\phi ({\bf x})\cos (\omega _{0}t),  \nonumber \\
\dot{q} &=&\phi ({\bf x})\sin (\omega _{0}t),
\end{eqnarray}
where $\phi ({\bf x})$ is any observable of the base dynamics and $\omega
_{0}\neq 0$ is an arbitrary constant frequency that is needed to damp off
any linear growth that may be common to both non-chaotic and chaotic orbits.
If the observable $\phi $ is drawn from a non-chaotic system, then the $(p,q)$
subspace will be bounded. But if the observable $\phi $ is drawn from
a chaotic system, then the motion in the $(p,q)$ subspace will be Brownian and
unbounded. The extended variables diffuse through the subspace as the
observables of the base dynamics jump around unpredictably.

Gottwald and Melbourne introduced a binary test for chaos by defining a
mean-square displacement 
\begin{equation}
M(t)=\lim_{T\rightarrow \infty }{\frac{1}{T}}\int_{0}^{T}\left( p(t+\tau
)-p(\tau )\right) ^{2}d\tau 
\end{equation}
and characterizing the behaviour of $M(t)$ through 
\begin{equation}
K=\lim_{t\rightarrow \infty }{\frac{\log M(t)}{\log t}}\ \ .
\end{equation}
If there is no chaos then the motion in $p(t)$ is bounded and 
$K\rightarrow 0$. If the base dynamics is chaotic then $p(t)$ will exhibit
Brownian diffusion so that $\Delta p\rightarrow t^{1/2}$ as 
$t\rightarrow \infty$ and
$K\rightarrow 1$.
The $K$-test can be used for both continuous and discrete dynamical systems
even when the precise underlying base dynamics is unknown \cite{gm}.

This test has certain advantages of simplicity and was shown to confirm the
results of testing for chaos using Lyapunov exponents for 
the forced van-der-Pol system \cite{gm}. We show here
that the $K$-test also confirms the trend of the Lyapunov exponents for the
simple Hamiltonian system of Henon and Heiles \cite{hh}. So the $K$-test is
applicable to Hamiltonian as well as dissipative systems.

While the test works well as a diagnostic of the transition from regularity
to chaos we express some reservations about the use of the test on an orbit
by orbit basis. Most importantly, we found the method difficult to interpret
for systems that experienced chaotic transients in contrast to the easy
interpretation of dissipative systems near a strange attractor, or in highly
chaotic Hamiltonian systems. To illustrate this we use the Lorenz model 
\cite{loref} to
show that transient chaos, which can be found in the Lorenz system by other
means, goes undetected by the $K$-test. Orbits that enter highly
chaotic regions of phase space and then depart into regular regions are
not Anosov and may not show the diffusive motion in phase space
required to yield a $K\rightarrow 1$ result from the test. 
These transient systems may
not satisfy the criteria for a system within the remit of the $K$-test. 
This signals one of the possible limitations of the K-test.

\section{Henon-Heiles}

The Henon-Heiles system is an ideal Hamiltonian system on which to evaluate
the utility of the $K$-test. It provides a simple and well understood model
for the motion of stars in a galactic potential as well as for the motion of
non-linearly-coupled molecules. The transition from regularity to chaos as
the energy is increased is well documented and can be identified with 
Poincar\'{e} 
surfaces of section, Lyapunov exponents, and the Painlev\'{e} property.

\begin{figure}
\centerline{\psfig{file=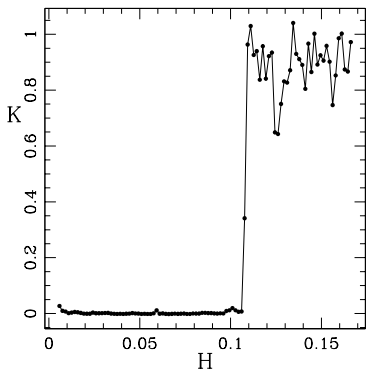,width=2.in}}
\vskip4pt
\caption{
The asymptotic value of $K$ versus the 
energy $H$ for the Henon-Heiles system.
\label{m}}  \end{figure}

\begin{figure}
\centerline{\psfig{file=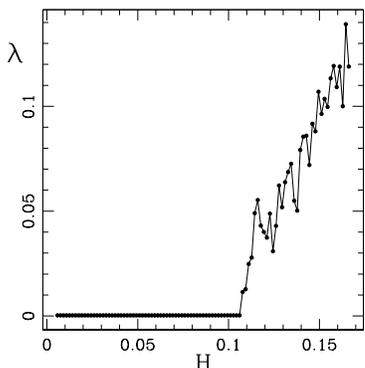,width=2.in}}
\caption{
The asymptotic value of the principal Lyapunov exponent 
versus the 
energy $H$ for the Henon-Heiles system.
\label{l}}  \end{figure}

The Henon-Heiles Hamiltonian is 
\begin{equation}
H={\frac{1}{2}}\left( p_{x}^{2}+p_{y}^{2}+x^{2}+y^{2}\right) +x^{2}y-
{\frac{1}{3}}y^{3}
\end{equation}
with equations of motion 
\begin{eqnarray}
\dot{x} &=&p_{x}  \nonumber \\
\dot{y} &=&p_{y}  \nonumber \\
\dot{p}_{x} &=&-x-2xy  \nonumber \\
\dot{p}_{y} &=&-y-x^{2}+y^{2}\ \ .  \label{henon}
\end{eqnarray}
We add the Euclidean extension 
\begin{eqnarray}
\dot{p} &=&x\cos (\theta )  \nonumber \\
\dot{q} &=&x\sin (\theta )  \nonumber \\
\dot{\theta} &=&\omega _{0}  \label{e2}
\end{eqnarray}
where we have chosen $\phi =x$. We evolve these equations numerically and
check that the Hamiltonian remains conserved throughout the simulation. Our
results are summarized in Figs.\ \ref{m} and \ref{l}. The orbits selected
for display have identical initial conditions up to the energy $H$ which was
varied from orbit to orbit. The value of $K(t)$ is measured as the slope
of the line $ln(M(t))$ versus $ln(t)$ and is
plotted in Fig.\ \ref{m}
as a function of the energy $H$. The transition from non-chaotic orbits 
($K\rightarrow 0 $) to chaotic orbits ($K\rightarrow 1$) as the energy 
increases is clearly
demonstrated. The transition confirms the break up of tori in the phase
space for the base dynamics (and therefore the non-integrability of the base
dynamics) as the energy grows. We also confirm the onset of chaotic motion
for these orbits with a numerical determination of the principal Lyapunov
exponent shown as a function of $H$ in Fig.\ \ref{l}.

A few comments on Fig.\ \ref{m} should be made. The convergence of the value
of $K$ towards either $0$ or $1$ can be improved if the numerical
simulations are allowed to run far longer and a more precise extrapolation
method is used so as to avoid any early transients. Of course, this renders
the application more numerically intensive and slower without yielding a
very great improvement in convergence. There is also a clear dependence of
the motion in the Euclidean subspace 
on the specific value of $\omega _{0}$ employed.
However, the overall trend in $K$ is 
typically the same for various values of $\omega _{0}$.

While Fig.\ \ref{m} does show the required transition from 0 to 1, it is 
important to stress how the conservative system differs from the dissipative
system. For a dissipative system all trajectories, regardless of initial
conditions, are drawn onto the same attractor. Therefore the $K$-test
should show no dependence on initial conditions. This is not true for a 
Hamiltonian system. There can be a mixture of regular and irregular
orbits and so the test result can depend on initial conditions. Only 
for a completely ergodic system will $K\rightarrow 1$ independent of 
initial conditions. Also, since chaotic transients can often arise in 
conservative systems, the $K$-test may not always give such a crisp 
transition. Chaotic transients in dissipative systems are discussed further
in the next section.

\section{Lorenz}

\begin{figure}
\centerline{\psfig{file=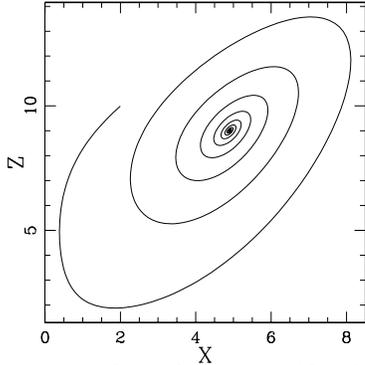,width=2.in}}
\caption{A nonchaotic orbit for $r=10$ being drawn onto the 
regular attracting fixed point at $(X_R,Y_R,Z_R)$.
The orbit began with
initial conditions $X(0)=1.99895$, $Y(0)=0$, and $Z(0)=9.99005$.
\label{simple}}  \end{figure}

The standard Lorenz system demonstrates a transition from no chaos to
transient chaos and then further to a full chaotic attractor. The standard
system is 
\begin{eqnarray}
\dot{X} &=&-\sigma (X-Y),  \nonumber \\
\dot{Y} &=&-XZ+rX-Y,  \nonumber \\
\dot{Z} &=&XY-bZ,
\end{eqnarray}
with $\sigma =10$ and $b=8/3$ and we vary the constant $r$. The transition
to chaos is summarized in Ref.\ \cite{ott}. For $r<1$ there is one fixed
point at $X=Y=Z=0$ and no chaos. For 
$1<r \mathrel{
\hbox to 0pt{\lower
3pt\hbox{$\mathchar"218$}\hss}\raise 2.0pt\hbox{$\mathchar"13C$}}13.96$
there are two additional attractors, one at 
\begin{eqnarray}
X_{R} &=&(b(r-1))^{1/2}  \nonumber \\
Y_{R} &=&(b(r-1))^{1/2}  \nonumber \\
Z_{R} &=&r-1,
\end{eqnarray}
and another at $(X_{L},Y_{L},Z_{L})=(-X_{R},-Y_{R},Z_{R}),$ but still no
chaos. A non-chaotic orbit drawn onto the simple fixed point 
$(X_{R},Y_{R},Z_{R})$ with $(\dot{X},\dot{Y},\dot{Z})\rightarrow 0$ is shown
in Fig.\ \ref{simple}. A closely neighbouring orbit will follow a similar
path onto the same attractor since there is no exponential sensitivity to
initial conditions, no positive Lyapunov exponent, and no chaos.

\begin{figure}
\centerline{\psfig{file=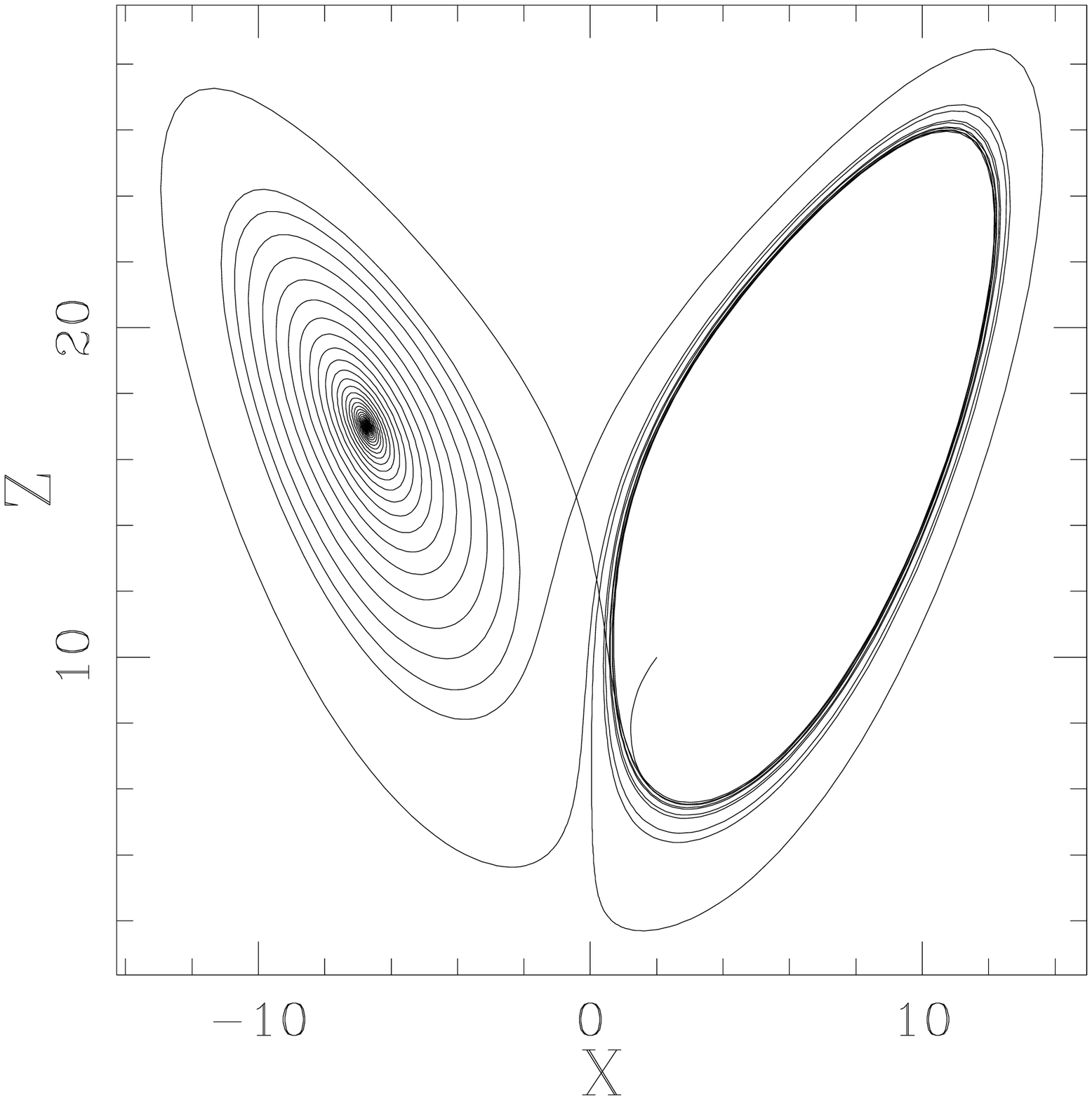,width=2.in}\psfig{file=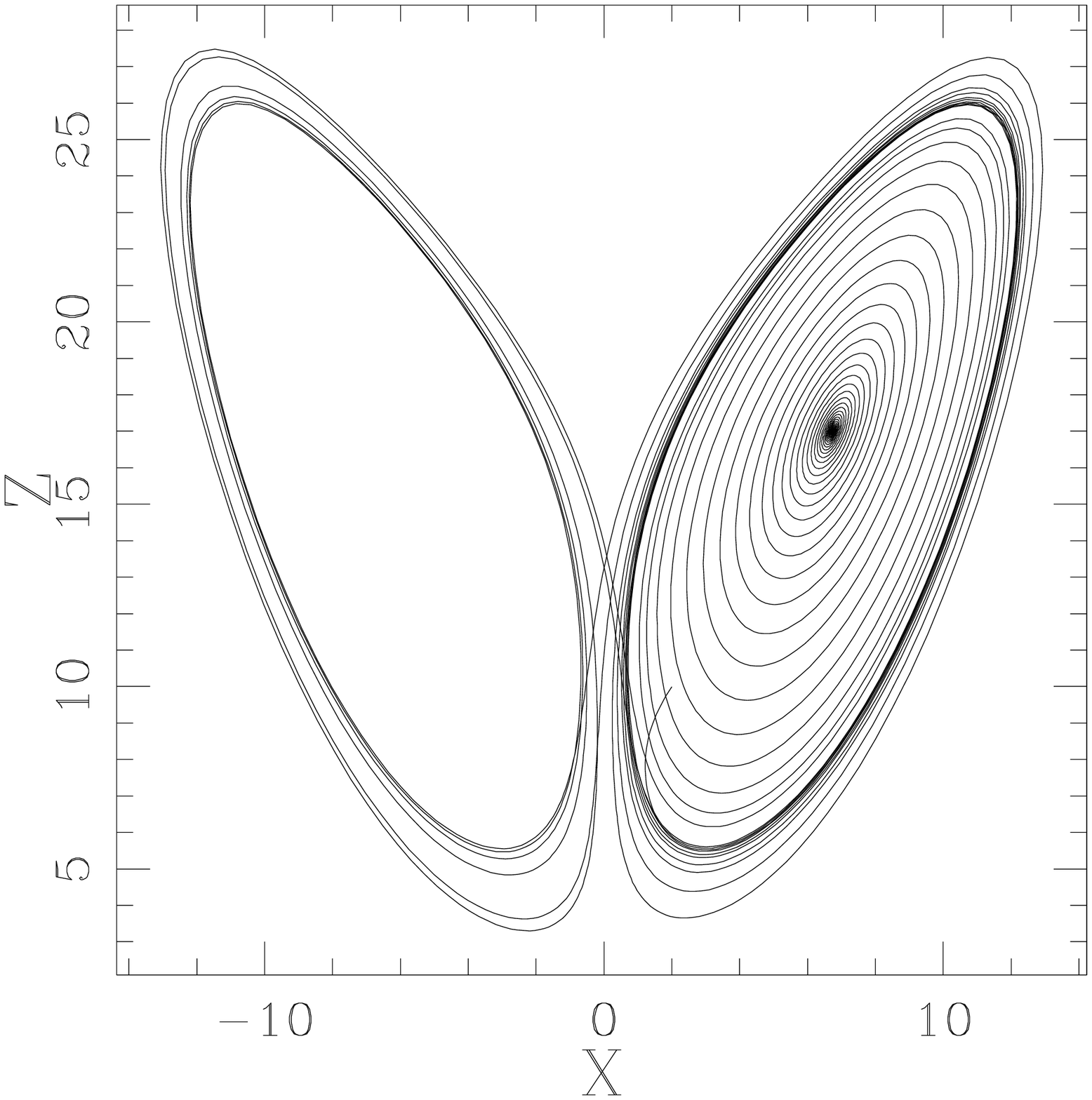,width=2.in}}
\caption{An orbit in the $(X,Z)$ plane
with $r=18$ exhibiting transient chaos before gliding 
towards the (not strange) attractor at $(X_L,Y_L,Z_L)$.
Left: The orbit began with
initial conditions $X(0)=1.99885$, $Y(0)=0$, and $Z(0)=9.99005$.
Right: A neighbouring orbit with
$X(0)=1.99895$, $Y(0)=0$, and $Z(0)=9.99005$ is drawn onto the other regular
attractor at $(X_R,Y_R,Z_R)$.
\label{transient}}  \end{figure}

However, for values of 
$r\mathrel{
\hbox to 0pt{\lower
3pt\hbox{$\mathchar"218$}\hss}\raise 2.0pt\hbox{$\mathchar"13E$}}13.96$ a
non-attracting form of transient chaos develops around a homoclinic orbit in
phase space, first noted by Kaplan and Yorke \cite{ky}. Two orbits
undergoing transient chaos are shown in Fig.\ \ref{transient}. One orbit
winds around a bit before being drawn onto the regular attracting point 
$(X_{L},Y_{L},Z_{L})$ while its very near neighbour ends up on 
$(X_{R},Y_{R},Z_{R})$.

\begin{figure}
\centerline{\psfig{file=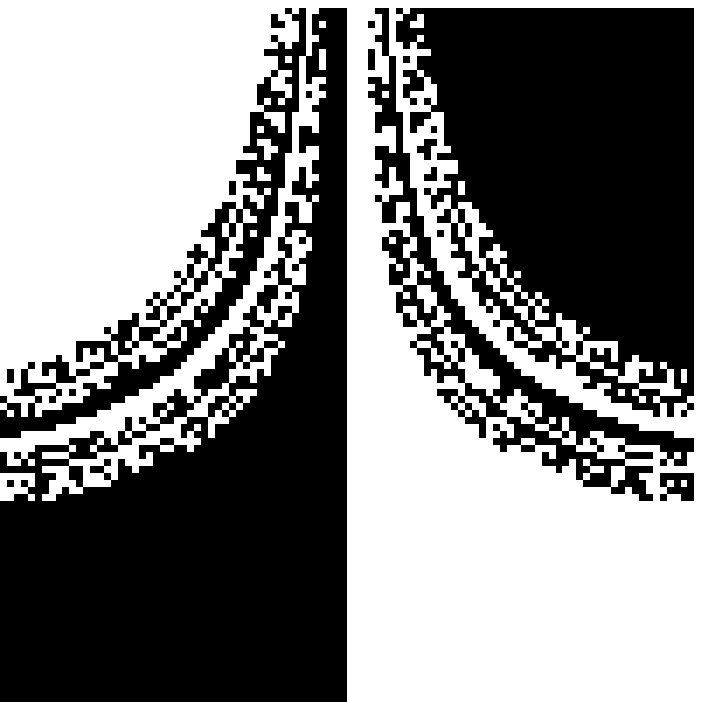,width=1.5in}}
\vskip 5truept
\centerline{\psfig{file=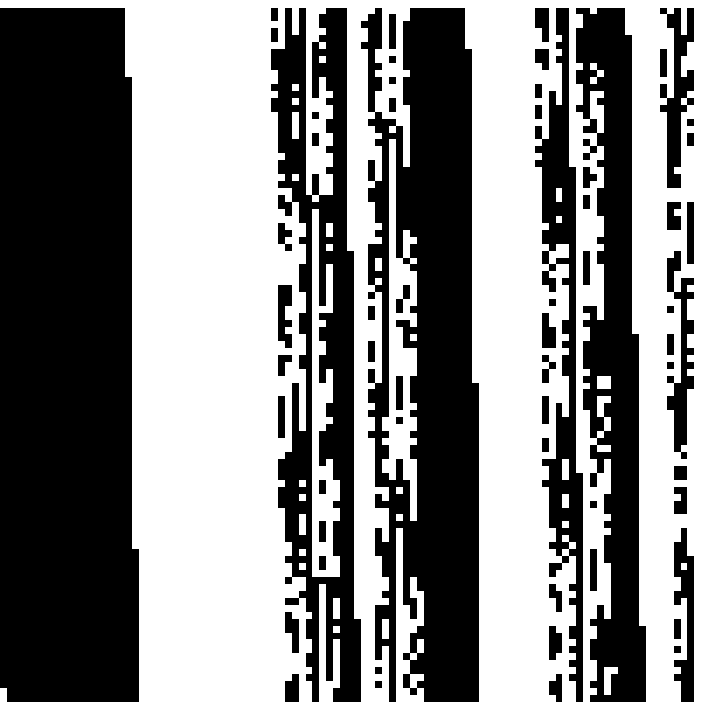,width=1.5in}}
\vskip 5truept
\centerline{\psfig{file=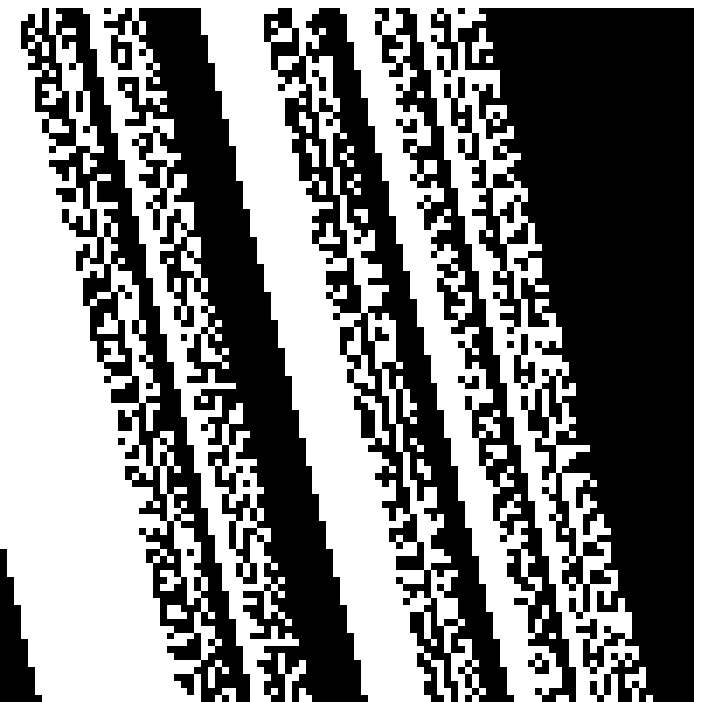,width=1.5in}}
\vskip 5truept
\caption{Upper Figure: A fractal basin boundary in $(X,Z)$ for $r=18$.
The range of initial conditions is $-10 \le X(0) \le 10$ and $0\le Z(0) \le 10$ while for all orbits $Y(0)=0$.
Middle Figure: A detail of the fractal basin boundary in $(X,Z)$.
The range of initial conditions is $1 \le X(0) \le 2$ and $9.9\le Z(0) \le 10$
while again $Y(0)=0$.
Lower Figure: A detail of the detail.
The range of initial conditions is $1.99 \le X(0) \le 2$ and $9.99\le Z(0) \le 10$ with $Y(0)=0$.
\label{fbb}}  \end{figure}

We choose to use the method of fractal basin boundaries to locate regions of
transient chaos. An initial slice in phase space is color-coded according to
whether the orbit eventually lands on the fixed point $(X_{R},Y_{R},Z_{R})$
(black) or the fixed point $(X_{L},Y_{L},Z_{L})$ (white). In the absence of
chaos all the basins will have smooth boundaries. In the presence of chaos
the boundaries become fractal, demonstrating both extreme sensitivity to
initial conditions and a chaotic mixing of orbits. Such fractal basin
boundaries are illustrated in Fig.\ \ref{fbb}; the orbits shown in Fig.\ 
\ref{transient} are drawn from along the fractal. They have nearly identical
initial values but divergent outcomes, a characteristic of chaotic
transients.

\begin{figure}
\centerline{\psfig{file=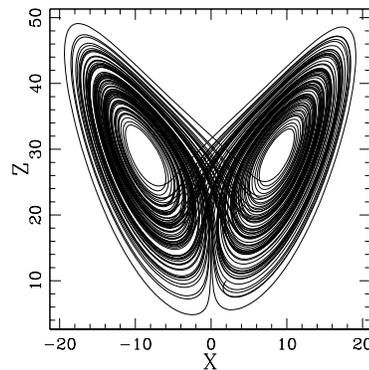,width=2.in}}
\caption{An orbit in the $(X,Z)$ plane along the strange attractor for $r=30$.
The orbit began with
initial conditions 
$X(0)=1.99895$, $Y(0)=0$, and $Z(0)=9.99005$.
\label{attractor}}  \end{figure} 


For $r
\mathrel{
\hbox to 0pt{\lower
3pt\hbox{$\mathchar"218$}\hss}\raise 2.0pt\hbox{$\mathchar"13E$}}24.06$ the
transient chaos gives way to chaos on attractors which merge into the famous
Lorenz strange attractor beyond $r
\mathrel{
\hbox to 0pt{\lower
3pt\hbox{$\mathchar"218$}\hss}\raise 2.0pt\hbox{$\mathchar"13E$}}24.74$ 
\cite{ott}. An orbit that drifts onto the strange Lorenz attractor is shown in
Fig.\ \ref{attractor}.

\begin{figure}
\centerline{\psfig{file=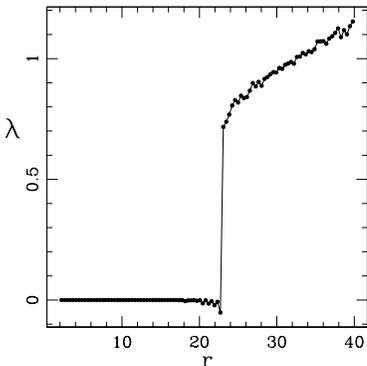,width=2.in}}
\caption{
The principal Lyapunov exponent versus the 
parameter $r$ for orbits with inital values 
$X(0)=1.99895$, $Y(0)=0$, and $Z(0)=9.99005$.
\label{lexplor}}  \end{figure}

We apply the $K$-test to the Lorenz system and show that it effectively
marks the transition from non-chaotic motion to chaos on a strange attractor
at $r\simeq 24.74$ where we see $K$ rising from $0$ to $1$. However, the
test is unable to pick up the chaotic transient behaviour for values of 
$13.96
\mathrel{
\hbox to 0pt{\lower
3pt\hbox{$\mathchar"218$}\hss}\raise 2.0pt\hbox{$\mathchar"13E$}}r
\mathrel{
\hbox to 0pt{\lower
3pt\hbox{$\mathchar"218$}\hss}\raise
2.0pt\hbox{$\mathchar"13E$}}24.06$.
Notice that a similar random scan through the Lyapunov exponents in 
Fig.\ \ref{lexplor} also misses the transient episodes although short 
time exponents along the fractal basin boundary can be isolated.

\begin{figure}
\centerline{\psfig{file=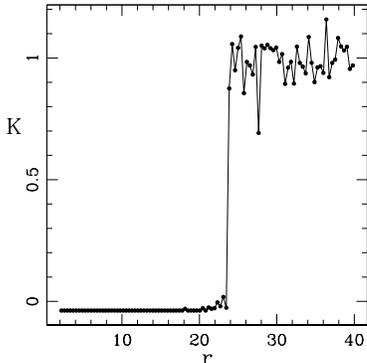,width=2.in}}
\caption{The asymptotic value of $K$ versus the 
parameter $r$ for orbits with inital values 
$X(0)=1.99895$, $Y(0)=0$, and $Z(0)=9.99005$.
\label{klor}}  \end{figure} 

\begin{figure}
\centerline{\psfig{file=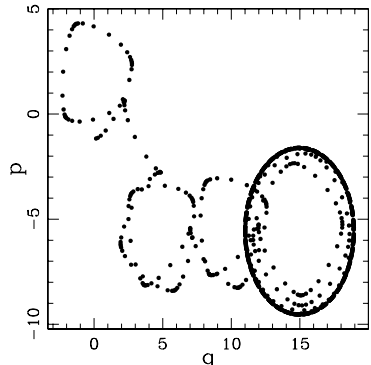,width=2.in}}
\caption{The bounded regularity of the 
$(p,q)$-extension
for an orbit sampled from Fig.\ \ref{klor}. 
The orbit corresponds to $r=18$ with inital values 
$X(0)=1.99895$, $Y(0)=0$, and $Z(0)=9.99005$
(see Fig.\ \ref{transient}).
\label{sample1}}  \end{figure}

Specific orbits of Fig.\ \ref{klor} can be isolated to illustrate the
behaviour in the $(p,q)$ subspace explicitly. The orbit at $r=18$
corresponds to the right-most orbit in Fig.\ \ref{transient}. The orbit
winds around before drifting onto the fixed point $(X_{R},Y_{R},Z_{R})$.
This is reflected in the $(p,q)$ subspace by the stray steps taken before
the orbit moves onto the smooth, bounded ring of the regular attractor as
shown in Fig.\ \ref{sample1}. This results in a regular oscillation in the
mean-square displacement $M(t)$ that will eventually die away to give 
$K\rightarrow 0,$ as shown in Fig.\ \ref{klor}. The $K$-test reflects the
regularity of the attractor and is unable to detect the subtle chaotic
transient in the early motion.

\begin{figure}
\centerline{\psfig{file=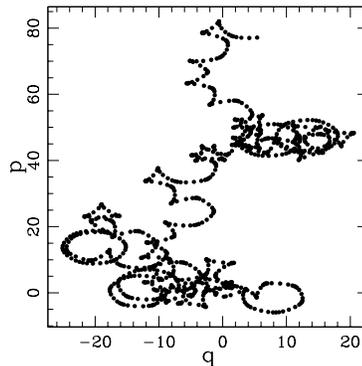,width=2.in}}
\caption{The figure shows the unbounded Brownian like diffusion of the 
$(p,q)$-extension
for a chaotic orbit for which $K\sim 1$ in Fig.\ \ref{klor}.
The orbit corresponds to $r=30$ with inital values 
$X(0)=1.99895$, $Y(0)=0$, and $Z(0)=9.99005$ and moves onto 
the Lorenz strange attractor.
\label{sample2}}  \end{figure}

By contrast, an orbit at $r=30$ in Fig.\ \ref{klor} is drawn onto the Lorenz
strange attractor and its strongly chaotic behaviour is detected by the 
$K$-test. The chaotic motion is well reflected by the Brownian diffusion
evidenced in the $(p,q)$ subspace shown in Fig.\ \ref{sample2}. This orbit
leads to $K\rightarrow 1$ as expected.

\section{Summary}

We have confirmed that the $K$-test provides a simple and easy diagnostic
for the transition from regularity to chaos for Hamiltonian as well as
dissipative systems. For the test to be effective, an orbit must spend
sufficient time on the hyperbolic chaotic attractor in a dissipative system,
or be confined to a highly chaotic region of phase space if the system is
Hamiltonian. However, for chaotic transients which move into a chaotic
region of phase space and then out into a regular region of phase space the
motion is not consistently Brownian. Consequently, the test can yield
ambiguous and confusing results although a qualitative look over the $(p,q)$
subspace can provide guidance as to the transient irregularity of the base
dynamics. Additionallly the K-test is not ideal for relativistic
settings since it depends on the time coordinate used and so, like the 
Lyapunov exponents, is not a covariant indicator of chaos.
These limitations of the $K$-test are not unexpected: no one probe
of chaos can suit every scenario. Nor are they fatal to its utility. Rather,
we hope that they will help to map out the territory over which this simple
test is a reliable guide to the presence of chaos

\section*{Acknowledgements}

We are especially grateful to both G. Gottwald and I. Melbourne for
answering questions regarding their method. JL thanks
N.J. Cornish for useful conversations. 
JL is supported by a PPARC Advanced Fellowship and an award from NESTA.

\end{document}